\documentclass[twocolumn,preprintnumbers,nofootinbib,
prd,aps]{revtex4}

\usepackage{epsf}
\usepackage{latexsym}
\usepackage{amssymb}
\usepackage{epsf}
\usepackage{amsmath}
\usepackage{slashed}

\def \pardif[#1][#2]{\frac{\partial #1}{\partial #2}}
\def \eqref[#1]{Eq.~(\ref{#1})}
\def \citeref[#1]{Ref.~\cite{#1}}
\def \figref[#1]{Fig.~\ref{#1}}
\def \secref[#1]{Sec.~\ref{#1}}

\begin{document}



\renewcommand{\topfraction}{0.8}

\preprint{UT-12-26}

\title{

Probing dark radiation with inflationary gravitational waves

}

\author{
Ryusuke Jinno, Takeo Moroi and Kazunori Nakayama
}

\affiliation{
Department of Physics, University of Tokyo, Bunkyo-ku, 
Tokyo 113-0033, Japan
}

\begin{abstract}

  Recent cosmological observations indicate the existence of extra
  light species, i.e., dark radiation.  In this paper we show that
  signatures of the dark radiation are imprinted in the spectrum of
  inflationary gravitational waves. If the dark radiation is
    produced by the decay of a massive particle, high frequency mode
    of the gravitational waves are suppressed.  In addition, due to
    the effect of the anisotropic stress caused by the dark radiation, a
    dip in the gravitational wave spectrum may show up at the
    frequency which enters the horizon at the time of the dark
    radiation production. Once the gravitational wave spectrum is
  experimentally studied in detail, we can infer the information on
  how and when the dark radiation was produced in the Universe.

\end{abstract}

\maketitle

\renewcommand{\thefootnote}{\#\arabic{footnote}}

\section{Introduction}

Recently, there are increasing evidence of the extra non-interacting
relativistic degrees of freedom, in addition to the standard three
(nearly) massless neutrino species.  The abundance of relativistic
component is parametrized by the effective number of neutrino species,
$N_{\rm eff}$, as
\begin{equation}
  \rho_{\rm rel}=\left[  
    1+N_{\rm eff}\frac{7}{8}\left(\frac{4}{11}\right)^{4/3} 
  \right] \rho_\gamma,
\end{equation}
where $\rho_{\rm rel}$ is the total relativistic energy density and
$\rho_\gamma = (\pi^2/15)T_\gamma^4$ denotes the photon energy density
measured after the $e^+e^-$ annihilation with $T_\gamma$ representing
the photon temperature.  The standard model predicts $N_{\rm
  eff}=3.05$~\cite{Mangano:2005cc}.

The $N_{\rm eff}$ can be constrained from various observations.
First, increasing $N_{\rm eff}$ leads to larger Hubble expansion rate
at the big bang nucleosynthesis epoch, which in turn results in
increase of the primordial helium abundance.  Recent observations
suggest $N_{\rm eff} = 3.68^{+0.80}_{-0.70}$ at $2\sigma$
level~\cite{Izotov:2010ca}.  (See, however, also
Ref.~\cite{Aver:2010wq} for discussion on the error estimation in the
helium abundance.)

The cosmic microwave background (CMB) anisotropy is also sensitive to
$N_{\rm eff}$.  The information on $N_{\rm eff}$ is imprinted in the
CMB anisotropy in some ways.  First, increase of $N_{\rm eff}$ makes
the early integrated Sachs-Wolfe effect more efficient, and the first
peak of the CMB power spectrum is enhanced.  Second, it tends to make
the scale of sound horizon smaller at the recombination epoch,
resulting in shift of the peak positions in the CMB power spectrum
toward high multipole moment.  Third, it erases the small scale power
spectrum due to the effect of free-streaming.  The WMAP seven-year
results combined with standard rulers give $N_{\rm eff} =
4.32^{+0.86}_{-0.88}$ at $1\sigma$ level~\cite{Komatsu:2010fb}.
Adding small scale CMB measurements improves the accuracy as $N_{\rm
  eff} = 4.56\pm 0.75$ for ACT~\cite{Dunkley:2010ge} and $N_{\rm eff}
= 3.86\pm 0.42$ for SPT~\cite{Keisler:2011aw} both at $1\sigma$ level.
Ref.~\cite{Archidiacono:2011gq} combined the WMAP, ACT and SPT
datasets with standard rulers and obtained $N_{\rm eff} =
4.08^{+0.71}_{-0.68}$ at $2\sigma$ level.\footnote{
	Note that the statistical significance depends on the prior on the Hubble parameter~\cite{Calabrese:2012vf}.
}
See also recent related studies~\cite{Hamann:2011hu,Nollett:2011aa,Hamann:2010bk,Hamann:2011ge}.

To summarize, at the current situation, observations suggest $\Delta
N_{\rm eff}\equiv N_{\rm eff}-3 \simeq 1$ at nearly $2\sigma$ level.
Motivated by these increasing evidence of the extra light species,
which is often called ``dark radiation'', models to explain dark
radiation were
proposed~\cite{Ichikawa:2007jv,Jaeckel:2008fi,Nakayama:2010vs,Fischler:2010xz,Kawasaki:2011ym,Hall:2011zq,Hasenkamp:2011em,Kawasaki:2011rc,Menestrina:2011mz,Kobayashi:2011hp,Hooper:2011aj,Jeong:2012hp,Kawakami:2012ke,Blennow:2012de,Moroi:2012vu}.
Although there are many candidates, if the dark radiation has only
extremely weak interaction with the standard model particles, it may
be difficult to detect it experimentally.  Thus it is important to
study how to confirm and distinguish models of dark radiation by other
observations.  For example, in
Refs.~\cite{Kawasaki:2011rc,Kawakami:2012ke} the possibility that the
dark radiation has (non-Gaussian) isocurvature perturbations was
considered. In Ref.~\cite{Zhao:2009aj} the effect of dark radiation on
CMB B-mode spectrum is discussed.

In this paper we consider a novel method to detect dark radiation
through inflationary gravitational waves (GWs).  It is known that
relativistic free streaming fluid can contribute to the anisotropic
stress, which potentially affects the propagation of
GWs~\cite{Weinberg:2003ur}.  This effect was concretely studied for
the free-streaming neutrinos.  GWs entering the horizon after the
neutrino freezeout dissipate their energies and, as a result, a
modulation feature shows up in the GW
spectrum~\cite{Weinberg:2003ur,Dicus:2005rh,Watanabe:2006qe}.  It was
also studied in the context of large lepton
asymmetry~\cite{Ichiki:2006rn}.  Therefore, it is expected that the
dark radiation also induces similar effects on GWs.  As opposed to the
case of neutrinos, we do not know when and how the dark radiation was
generated in the Universe.  Thus the position and strength of the
modulation in the GWs, if detected, tells us exactly about the
production mechanism of dark radiation.  In particular, models of dark
radiation produced by decay of non-relativistic fields
\cite{Ichikawa:2007jv,Fischler:2010xz,Kawasaki:2011ym,Hasenkamp:2011em,Kawasaki:2011rc,Menestrina:2011mz,Kobayashi:2011hp,Hooper:2011aj,Jeong:2012hp}
shows characteristic features in the primordial GW spectrum.  The
feature consists of combination of the anisotropic stress effect and
the modified background expansion history.  This is detectable in
future space-based GW detectors such as DECIGO~\cite{Seto:2001qf} and
BBO~\cite{Crowder:2005nr,Cutler:2009qv}.  We also show that the GW
spectrum will be a powerful tool for confirming the dark radiation
produced thermally and decoupled at some epoch in the early Universe.

This paper is organized as follows.  In Sec.~\ref{sec:DR} we review a
model of dark radiation produced by decaying particles.  In
Sec.~\ref{sec:GW} we calculate the evolution of gravitational waves in
the presence of anisotropic stress induced by dark radiation, and show
that characteristic signatures appear in the spectrum.
Sec.~\ref{sec:conc} is devoted to conclusions and discussion.

\section{Dark radiation production by decaying particles}   \label{sec:DR}

\subsection{Background evolution} 

We consider the case where the non-relativistic matter $\phi$ decays
into $X$ particle which plays the role of dark radiation. Thus $X$ is
assumed to be massless and has no interaction with other fields.  To
be more precise, $X$ must be relativistic until the recombination
epoch and its interaction must be so weak that remains to be decoupled
from thermal bath after the production by $\phi$ decay.  The evolution
equations of components are given by
\begin{eqnarray}
  && \dot\rho_\phi + 3H \rho_\phi = -\Gamma_\phi \rho_\phi, 
  \label{eq_BG1}\\  
  && \dot\rho_{\rm rad} + 4H\rho_{\rm rad} = \Gamma_\phi(1-B_X)\rho_\phi,\\
  && \dot\rho_X + 4H \rho_X = \Gamma_\phi B_X \rho_\phi, 
  \label{eq_BG3}
\end{eqnarray}
where the dot represents time derivative, and the Friedmann equation,
\begin{equation}
  H^2 = \frac{\rho_{\rm tot}}{3M_P^2}
  =\frac{\rho_\phi + \rho_{\rm rad} + \rho_X}{3M_P^2},
  \label{eq_H}
\end{equation}
where $\rho_\phi, \rho_{\rm rad}$ and $\rho_X$ are energy densities of
$\phi$, visible radiation and dark radiation, respectively, $M_P$ is
the reduced Planck scale, $\Gamma_\phi$ is the decay rate of $\phi$,
and $B_X$ denotes its branching fraction into $X$.

The extra effective number of neutrino species is given by
\begin{equation}
  \Delta N_{\rm eff} = 
  \frac{43}{7}\left[ \frac{10.75}{g_{*s}(T_\phi)} \right]^{1/3} 
  \left[\frac{\rho_X}{\rho_{\rm rad}}\right]_{H\ll\Gamma_\phi},
  \label{DeltaN}
\end{equation}
where $g_{*s}(T_\phi)$ denotes the relativistic degrees of freedom at
$T=T_\phi$ where the $\phi$ decays, and $\rho_X$ and $\rho_{\rm rad}$
are evaluated well after the $\phi$ decay.  In our numerical study we
take the standard-model value of $g_{*s}(T_\phi)=106.75$, because as
we will see, $T_\phi \gg T_{\rm EW} \simeq \mathcal O(100)\,{\rm GeV}$
is necessary for observation.

\begin{figure}[t]
  \begin{center}
    \centerline{\epsfxsize=0.45\textwidth\epsfbox{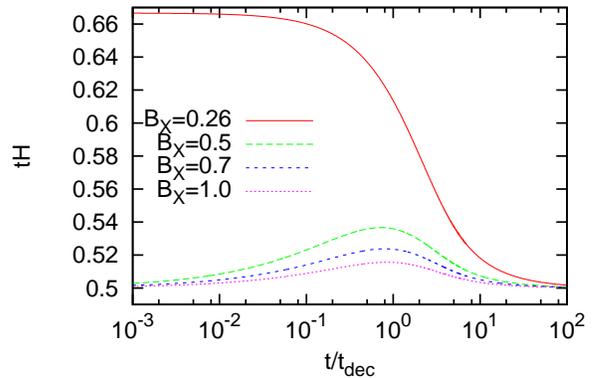}}
    \caption{Evolution of the product $tH$ as a function of cosmic
      time (normalized by $t_{\rm dec}$) for $B_X=0.26$ (red solid),
      $0.5$ (green dashed), $0.7$ (blue dotted) and $1.0$ (magenta
      dot-dashed) for explaining $\Delta N_{\rm eff} = 1$.}
    \label{fig:t_tH}
  \end{center}
\end{figure}

In order to obtain $\Delta N_{\rm eff}\simeq 1$, the energy density of
$\phi$ should nearly dominate the Universe at the decay.  Therefore,
the expansion rate of the Universe around the $\phi$ decay epoch is
modified.  Fig.~\ref{fig:t_tH} shows the product $tH$ as a function of
cosmic time $t$ normalized by $t_{\rm dec}$, defined by 
\begin{equation}
  t_{\rm dec} \equiv \frac{1}{\Gamma_{\phi}}.
\end{equation}
Here we have fixed initial conditions of $\rho_\phi$ and $\rho_{\rm
  rad}$ so that $\Delta N_{\rm eff}=1$ is realized.  Solid (red),
long-dashed (green), short-dashed (blue) and dotted (magenta) lines
correspond to $B_X=0.26$, $0.5$, $0.7$ and $1.0$, respectively.  It is
seen that $\phi$ has significant energy fraction around its decay and
the expansion rate is modified from the radiation-dominated one
($tH=1/2$).  In the limit of $\phi$ domination, we need $B_X \simeq
0.26$ for $\Delta N_{\rm eff} = 1$.  Thus $B_X \gtrsim 0.26$ is
required in order to realize $\Delta N_{\rm eff} = 1$: otherwise, the
$X$ energy density would be too small to explain $\Delta N_{\rm eff} =
1$ even if $\phi$ dominates the Universe.  Since the background
expansion rate is imprinted in the GW
spectrum~\cite{Turner:1990rc,Seto:2003kc,Tashiro:2003qp,Boyle:2005se,Boyle:2007zx,Nakayama:2008ip,Nakayama:2008wy,Kuroyanagi:2008ye,Mukohyama:2009zs,Nakayama:2009ce,Nakayama:2010kt,Schettler:2010dp,Durrer:2011bi,Kuroyanagi:2011fy,Jinno:2011sw,Saito:2012bb},
a particular shape in the GW spectrum is expected if dark radiation is
produced by decaying matter, as we will see.

\subsection{Model} 

As one of the motivated models of $\phi$ and $X$, we consider the
saxion and axion in a supersymmetric axion
model~\cite{Rajagopal:1990yx}.  This possibility was studied in
Refs.~\cite{Ichikawa:2007jv,Kawasaki:2011ym,Kawasaki:2011rc,Jeong:2012hp,Moroi:2012vu}
in the context of dark radiation.

The axion is a pseudo Nambu-Goldstone boson associated with the
spontaneous breakdown of the global U(1)$_{\rm PQ}$
symmetry~\cite{Peccei:1977hh}.  It solves the strong CP problem in the
quantum chromodynamics.  The axion has interactions suppressed by the
U(1)$_{\rm PQ}$ breaking scale, $f_a$.  The value of $f_a$ is
phenomenologically constrained as $10^9$\,GeV $\lesssim f_a \lesssim
10^{12}$\,GeV, and the axion mass is $\sim 10^{-2}$--$10^{-5}$\,eV for this
range of $f_a$~\cite{Kim:1986ax}.  Thus the axion is a good candidate
of dark radiation.

In a supersymmetric extension of the axion model, there appears a
scalar partner of the axion, called saxion, which is massless in
supersymmetric limit but obtains a mass from supersymmetry breaking
effects.  Writing the saxion mass as $m_\phi$, the saxion decay rate
into the axion pair is given by~\cite{Chun:1995hc}
\begin{equation}
  \Gamma_\phi = \frac{\xi^2}{64\pi}\frac{m_\phi^3}{f_a^2}.
\end{equation}
where $\xi$ is a model-dependent constant of order unity.  Assuming
that the saxion decays in the radiation dominated era in order to make
the signal detectable, the temperature at the saxion decay is
estimated to be
\begin{equation}
  T_\phi \sim 
  3\times 10^6\,{\rm GeV}\left( \frac{m_\phi}{10^3\,{\rm TeV}} \right)^{3/2}
  \left( \frac{10^{10}\,{\rm GeV}}{f_a} \right).
  \label{Tphi}
\end{equation}
The saxion with mass of $\mathcal O(10^3)\,{\rm TeV}$ is plausible by
taking account of the preference for high-supersymmetry breaking
scale~\cite{Giudice:2011cg} in light of the recent discovery of the
Higgs boson mass of 125\,GeV~\cite{Higgs}.  The saxion often
dominantly decays into the axion pair $(B_X\simeq 1)$.  The produced
axions are never thermalized below the temperature $\sim 10^7$\,GeV
for $f_a \gtrsim 10^{10}$\,GeV~\cite{Graf:2010tv}.  Assuming that the
saxion begins a coherent oscilation at $H=m_\phi$ with initial
amplitude of $\phi_i$, the saxion abundance in terms of the
energy-to-entropy ratio is given by $\rho_\phi/s\sim T_{\rm
  R}(\phi_i/M_P)^2$, where $T_{\rm R}$ is the reheating temperature
after inflation.  Then the abundance of relativistic axion after the
$\phi$ decay is estimated to be
\begin{equation}
  \left[\frac{\rho_X}{\rho_{\rm rad}}\right]_{H\ll\Gamma_\phi}
  \sim B_X\left[\frac{\rho_\phi}{\rho_{\rm tot}}\right]_{H=\Gamma_\phi} 
  \simeq \frac{B_X}{6}\frac{T_{\rm R}}{T_\phi}\left(\frac{\phi_i}{M_P}\right)^2.
\end{equation}
Therefore, for appropriate choices of $T_{\rm R}$ and $\phi_i$, e.g.,
for $T_{\rm R} \sim T_{\phi}$ and $\phi_i\sim M_P$, the axion
abundance produced by the saxion decay can account for the dark
radiation : $\Delta N_{\rm eff}\simeq 1$ (see Eq.~(\ref{DeltaN})).

\section{Spectrum of gravitational wave background with dark radiation}   \label{sec:GW}

\subsection{Evolution equations} 

Now let us study the evolution of primordial GWs under the presence of
dark radiation.  The GW corresponds to the tensor perturbation of the
metric.  We define the line element as
\begin{equation}
  ds^2 =-dt^2+a^2(t)(\delta_{ij}+h_{ij})dx^idx^j,
  \label{eq_line_element}
\end{equation}
where $h_{ij}$ is the transverse and traceless part of the metric
perturbation, and the Fourier amplitude of $h_{ij}$ as
\begin{eqnarray}
\begin{split}
  h_{ij}(t,\mathbf{x})
  &=
  \int \frac{d^3k}{(2\pi)^3}
  h_{ij} (t,\mathbf{k}) e^{i \mathbf{k} \mathbf{x}} \\
  &=
  \sum_{\lambda=+,\times} \int \frac{d^3k}{(2\pi)^3}
  h^{(\lambda)} (t,\mathbf{k})
  \epsilon_{ij}^{(\lambda)} e^{i \mathbf{k} \mathbf{x}},
  \end{split}
  \label{eq_h_decompose}
\end{eqnarray}
where $\epsilon_{ij}^{(\lambda)}$ denotes the polarization tensor.
As shown in Appendix, $h^{(\lambda)} (t,\mathbf{k})$ satisfies the
following equation
\begin{equation}
\begin{split}
  \ddot{h}^{(\lambda)} &(t,\mathbf{k})
  + 3 H \dot{h}^{(\lambda)} (t,\mathbf{k})
  + \frac{k^2}{a^2} h^{(\lambda)} (t,\mathbf{k}) \\
  = 
  &- 24 H^2 \frac{1}{a^4 (t) \rho_{\rm tot} (t) } \\
  &\times
  \int_0^t a^4 (t') \rho_X (t')
  K \left (k\int_{t'}^t \frac{dt''}{a(t'')} \right)
  \dot{h}^{(\lambda)} (t',\mathbf{k}) dt', \\
  \label{eq_hdif_t}
  \end{split}
\end{equation}
where 
\begin{equation}
  K(u) \equiv \frac{j_2(u)}{u^2} = - \frac{\sin(u)}{u^3} 
  - \frac{3\cos(u)}{u^4} + \frac{3\sin(u)}{u^5},
\end{equation}
with $j_2$ being the second-order spherical Bessel function.  Here we
have assumed that there is no source for the anisotropic stress except
for that induced by GW effects on dark radiation.  Contrary to the
case of neutrinos studied in
Refs.~\cite{Weinberg:2003ur,Watanabe:2006qe}, $\rho_X(t')$ is inside
the time integral since $\rho_X$ does not scale as $a^{-4}$ while $X$
is produced by the $\phi$ decay.  In terms of $u$ and $u'$ defined as
\begin{eqnarray}
  u &=& k \eta = k \int_0^t \frac{dt'}{a(t')},
  \\
  u' &=& k \eta' = k \int_0^{t'} \frac{dt''}{a(t'')},
\end{eqnarray}
where $\eta \equiv \int_0^t \frac{dt'}{a(t')}$ is the conformal time,
Eq.~(\ref{eq_hdif_t}) becomes
\begin{equation}
\begin{split}
  \pardif[^2 h^{(\lambda)}][u^2] &(u,\mathbf{k})
  + 2 H_u \pardif[h^{(\lambda)}][u] (u,\mathbf{k})
  + h^{(\lambda)} (u,\mathbf{k}) \\
  = 
  &-24 H_u^2 \frac{1}{a^4 (u) \rho_{\rm tot} (u) } \\
  &\times
  \int_0^u a^4 (u') \rho_X (u') K (u-u')
  \pardif[h^{(\lambda)}][u] (u',\mathbf{k}) du',
  \label{eq_hdif_u}
\end{split}
\end{equation}
with $H_u \equiv \frac{1}{a} \pardif[a][u]$. The RHS of \eqref[eq_hdif_u] 
have effects mainly for $u \sim 1$, which roughly equals to the time of 
horizon-crossing, $k=aH$. 


We have solved Eq.~(\ref{eq_hdif_u}) together with the
background evolution (\ref{eq_BG1}) -- (\ref{eq_BG3}) to derive the
present GW spectrum.

\subsection{Overall normalization} 

Before showing the detailed results, we here comment on the
normalization of the present GW energy density.  During inflation,
quantum fluctuations of the tensor perturbation is continuously
generated which turn into stochastic GW background in the present
Universe after the horizon-in~\cite{Maggiore:1999vm}.  It predicts
nearly scale invariant GW spectrum for the GW modes entering in the
horizon in the radiation-dominated
era~\cite{Allen:1987bk,Sahni:1990tx,Turner:1993vb,Turner:1996ck,Smith:2005mm,Smith:2006xf,Chongchitnan:2006pe,Friedman:2006zt}.
The GW energy density per log frequency at the horizon crossing
$k=aH$, normalized by the critical energy density, is given
by~\cite{Smith:2008pf}
\begin{eqnarray}
  \Omega_{\rm GW}(k=aH)=\frac{\Delta_h^2(k)}{24}
  \simeq\frac{2.43\times 10^{-9}r}{24}\left( \frac{k}{k_0} \right)^{n_t},
\end{eqnarray}
where $r$ denotes the tensor-to-scalar ratio, $n_t$ is the tensor
spectral index, $k_0 = 0.002\,{\rm Mpc^{-1}}$ is the pivot scale and
\begin{eqnarray}
  \Delta_h^2(k) \equiv 
  \frac{8}{M_P^2}\left( \frac{H_{\rm inf}}{2\pi} \right)^2 
  \left( \frac{k}{k_0} \right)^{n_t},
\end{eqnarray}
with $H_{\rm inf}$ being the Hubble scale during inflation and we have
assumed the WMAP normalization on the curvature perturbation on large
scale~\cite{Komatsu:2010fb}.  In this subsection, we consider the modes which
enter the horizon in the radiation-dominated era, since we are
interested in the high-frequency GWs which may be observed by
space-based GW detectors.

First, in the standard model without dark radiation, the present
spectrum of GWs is given by
\begin{eqnarray}
  \Omega_{\rm GW}^{\rm (SM)} (k) = \gamma^{\rm (SM)} \Omega_{\rm rad}^{\rm (SM)}\times  \Omega_{\rm GW}(k=aH),
  \label{eq_OmegaSM}
\end{eqnarray}
where $\Omega_{\rm rad}^{\rm (SM)}=4.2\times 10^{-5} h^{-2}$ with $h$
parameterizing the present Hubble parameter as $H_0=100h$\,km/s/Mpc
and
\begin{equation}
	\gamma^{\rm (SM)} = \left[\frac{g_*(T_{\rm in}(k))}{g_{*0}^{\rm (SM)}} \right]
	\left[\frac{g_{*s0}^{\rm (SM)}}{g_{*s}(T_{\rm in}(k))} \right]^{4/3},
\end{equation}
where $g_{*0}^{\rm (SM)} = 3.36$ and $g_{*s0}^{\rm (SM)} = 3.91$, and
$T_{\rm in}(k)$ denotes the temperature at which the mode $k$ enters
the horizon. \eqref[eq_OmegaSM] reflects the fact that GWs behave as a
relativistic component after they have entered the horizon.  We have
$\gamma^{\rm (SM)} \simeq 0.39$ for $g_*(T_{\rm in}(k)) = 106.75$.
The present GW spectrum per log frequency is then given by
\begin{eqnarray}
  \Omega_{\rm GW}^{\rm (SM)} (k) &\simeq& 3.3\times 10^{-16} 
  \nonumber \\ && \times
  \left( \frac{r}{0.1} \right)
  \left( \frac{k}{k_0} \right)^{n_t}
  \left[ \frac{106.75}{g_{*}(T_{\rm in}(k))} \right]^{1/3}.
\end{eqnarray}

In the presence of dark radiation, the overall normalization of the GW
spectrum is modified due to the change of expansion rate.  Neglecting
the effect of anisotropic stress, we find
\begin{eqnarray}
   \Omega_{\rm GW} (k) = \gamma \Omega_{\rm rad}\times  \Omega_{\rm GW}(k=aH),
\end{eqnarray}
where $\Omega_{\rm rad} = \Omega_{\rm rad}^{\rm (SM)}\times
(g_{*0}/g_{*0}^{\rm (SM)})$ with
\begin{equation}
  g_{*0} = 2\left[ 1+N_{\rm eff} \frac{7}{8}\left( \frac{4}{11} \right)^{4/3} \right].
\end{equation}
We find $g_{*0}\simeq 3.82$ for $N_{\rm eff}=4$. The factor $\gamma$
is given by
\begin{equation}
  \gamma = \frac{1+\frac{7}{43}\left( \frac{g_{*s}(T_\phi)}{10.75}  \right)^{1/3}\Delta N_{\rm eff} }
  {1/\gamma^{\rm (SM)}+\frac{7}{43}\left( \frac{g_{*s}(T_\phi)}{10.75}\right)^{1/3} \Delta N_{\rm eff}},
\end{equation}
where we have used the relation (\ref{DeltaN}).  Therefore, the
overall enhancement factor for the GW spectrum is given by
\begin{equation}
  C_1\equiv \frac{\gamma}{\gamma^{\rm (SM)}}
  \frac{g_{*0}}{g_{*0}^{\rm (SM)}},
  \label{normC1}
\end{equation}
which is $1.35$ for $\Delta N_{\rm eff}=1$.  The first factor comes
from the modified expansion rate between the horizon-in and
matter-radiation equality, and the second factor comes from the change
of the epoch of matter-radiation equality.  Thus, without the effect
of anisotropic stress, the GW amplitudes at high frequencies inferred
from the measured tensor-to-scalar ratio at the CMB scales are
enhanced in the presence of dark radiation.

\begin{figure}[t]
  \begin{center}
   \centerline{\epsfxsize=0.45\textwidth\epsfbox{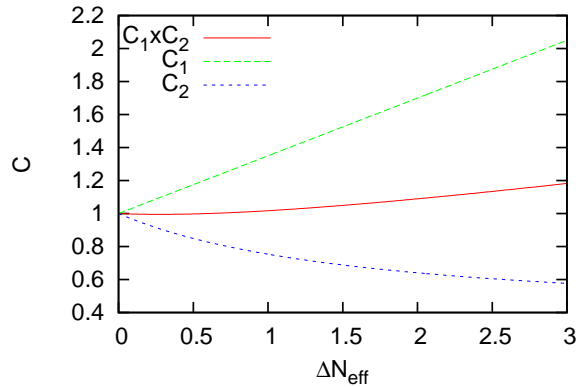}}
   \caption{ Relative normalization of the GW spectrum for $k_{\rm EW}
     \ll k \ll k_{\rm dec}$ as a function of $\Delta N_{\rm eff}$.
     $C_1$ (green dashed) : the enhancement factor due to the modified
     background evolution.  $C_2$ (blue dotted) : the suppression
     factor due to the anisotropic stress caused by dark radiation.
     Their product $C_1\times C_2$ (red solid) determines the
     resulting overall normalization for the modes $k_{\rm EW} \ll k
     \ll k_{\rm dec}$.  }
    \label{fig:norm}
  \end{center}
\end{figure}

Such an enhancement is compensated by the dissipation of GWs caused by
the anisotropic stress of dark radiation.  The suppression factor due
to the anisotropic stress, which we express here by $C_2$, was
analytically derived in Ref.~\cite{Dicus:2005rh,Boyle:2005se} as a
function of energy fraction of relativistic free-streaming particles
with respect to the total radiation energy density, which was assumed
to be constant around the time of horizon-crossing.\footnote{ See
    Eq.~(66) of Ref.~\cite{Boyle:2005se}.  Note that $C_3$ in
    Ref.~\cite{Boyle:2005se} corresponds to our $C_2$.  }  Thus we can
  apply their result to the present situation only for $k \ll k_{\rm
    dec}$, where $k_{\rm dec}$ denotes the comoving Hubble scale at
  $t=t_{\rm dec}$ :
\begin{eqnarray}
  k_{\rm dec} \equiv a(t_{\rm dec}) H(t_{\rm dec}),
\end{eqnarray}
because $\phi$ has completely decayed and the energy fraction of $X$ is constant
after the horizon crossing for the mode $k \ll k_{\rm dec}$.
In terms of $T_\phi$, it is given by
\begin{equation}
	k_{\rm dec} \simeq 0.27\,{\rm Hz} \left( \frac{T_\phi}{10^7\,{\rm GeV}} \right)
	\left( \frac{g_{*s}(T_\phi)}{106.75} \right)^{1/6}.
\end{equation}
If it is around $\mathcal O(1)$\,Hz, the GW features around $k\sim
k_{\rm dec}$ is observable at
DECIGO/BBO~\cite{Seto:2001qf,Crowder:2005nr}, which we will see in the
next subsection.  Thus we need $T_\phi \sim 10^6$--$10^8$\,GeV for
successful observation, which is actually the case for some particle
physics models, e.g., the saxion model (see Eq.~(\ref{Tphi})). 

The relative normalization for the GW spectrum is then given by the
product of them,
\begin{equation}
  \frac{\Omega_{\rm GW} (k)}{\Omega_{\rm GW}^{\rm (SM)} (k)} = C_1\times C_2.
  \label{norm}
\end{equation}
Fig.~\ref{fig:norm} shows $C_1$, $C_2$ and their product as functions
of $\Delta N_{\rm eff}$ for $k_{\rm EW} \ll k \ll k_{\rm dec}$ where
$k_{\rm EW}$ denotes the comoving Hubble scale around the electroweak
phase transition.  
(Note that $C_1$ depends on $k$ through $g_*(T_{\rm in}(k))$. 
For $k<k_{\rm EW}$, the value of $C_1$ is slightly modified. )
It is seen that there is a cancellation between
$C_1$ and $C_2$, and the result is close to one for $\Delta N_{\rm
  eff}=\mathcal O(1)$.  Although Eq.~(\ref{norm}) gives normalization
of the GW spectrum for $k_{\rm EW} \ll k \ll k_{\rm dec}$, the precise
shape of the GW spectrum around $k\sim k_{\rm dec}$ needs to be
investigated numerically.  Detailed results are shown in the next
subsection.

\subsection{Results} 

\begin{figure}[t]
  \begin{center}
    \centerline{\epsfxsize=0.45\textwidth\epsfbox{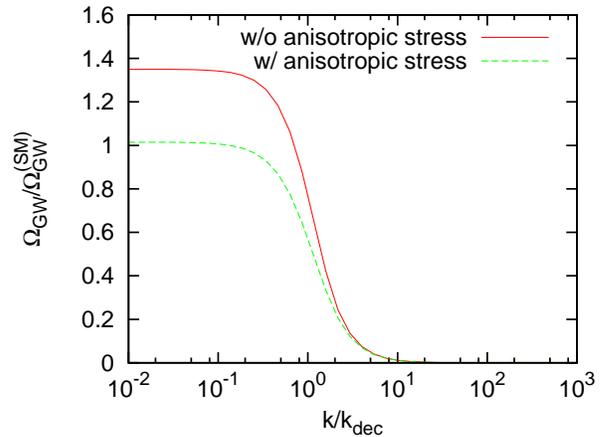}}
    \caption{$\Omega_{\rm GW}(k)/\Omega_{\rm GW}^{\rm (SM)}(k)$ as a
      function of $k$ (normalized by $k_{\rm dec}$) for $B_X=0.26$.
      The green (dotted) line is the full result, taking account of
      the effect of anisotropic stress.  For comparison, in the solid
      line (red), we also plot $\Omega_{\rm GW}(k)/\Omega_{\rm
        GW}^{\rm (SM)}(k)$ without the effect of anisotropic stress.
      The value at the plateau for $k\ll k_{\rm dec}$ is given by $C_1
      \times C_2$ and $C_1$ with and without anisotropic stress,
      respectively.  }
    \label{fig:k_omega_0.259207}
  \end{center}
\end{figure}

\begin{figure}[t]
  \begin{center}
   \centerline{\epsfxsize=0.45\textwidth\epsfbox{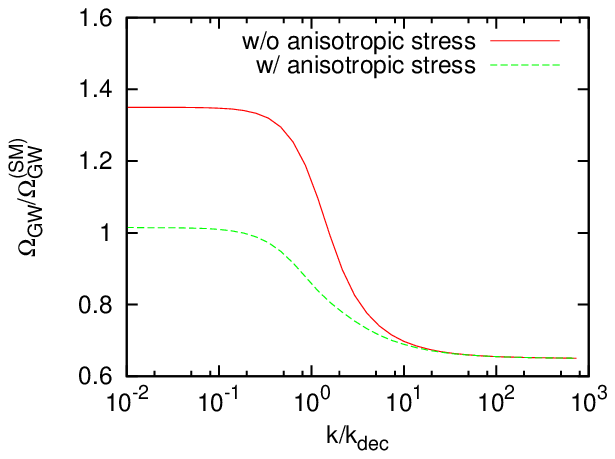}}
    \caption{Same as Fig.\ \ref{fig:k_omega_0.259207} but for $B_X = 0.5$.}
    \label{fig:k_omega_0.5}
  \end{center}
\end{figure}

\begin{figure}[t]
  \begin{center}
   \centerline{\epsfxsize=0.45\textwidth\epsfbox{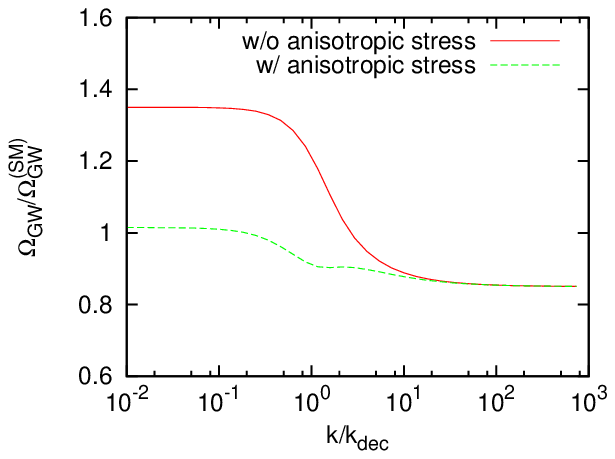}}
    \caption{Same as Fig.\ \ref{fig:k_omega_0.259207} but for $B_X = 0.7$.}
    \label{fig:k_omega_0.7}
  \end{center}
\end{figure}

\begin{figure}[t]
  \begin{center}
   \centerline{\epsfxsize=0.45\textwidth\epsfbox{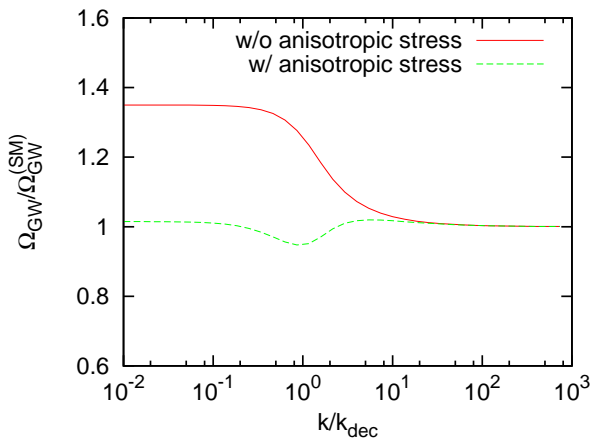}}
    \caption{Same as Fig.\ \ref{fig:k_omega_0.259207} but for $B_X = 1.0$.}
    \label{fig:k_omega_1.0}
  \end{center}
\end{figure}

In Figs.\ \ref{fig:k_omega_0.259207} -- \ref{fig:k_omega_1.0}, we plot
the GW spectrum normalized by $\Omega_{\rm GW}^{\rm (SM)} (k)$ predicted in the present scenario, 
varying $B_X$ from 0.26 to 1.0.  
The horizontal axis is normalized by $k_{\rm dec}$.
For comparison, we have also plotted the GW spectrum without
the effect of anisotropic stress. 
As one can see, the spectrum of the GWs has a characteristic change at
$k\sim k_{\rm dec}$ if the dark radiation (with $\Delta N_{\rm eff}\sim
1$) is produced by the decay of massive particle.  Thus, once the GW
spectrum is precisely measured, we have a chance to extract the
information on the mechanism of dark-radiation production.


There are several effects on the GW spectrum in the presence of dark
radiation.  First, since $\phi$ (nearly) dominates the Universe at the
time of its decay in order to realize $\Delta N_{\rm eff}\sim 1$,
$\Omega_{\rm GW}$ decreases at $k\gtrsim k_{\rm dec}$.  This is due to
the change of equation of state of the Universe.  The GW energy
density scales as $a^{-4}$ inside the horizon, while total energy
density scales as $a^{-3}$ in the $\phi$-dominated period.  Even if
$\phi$ does not completely dominate the Universe, there should be
deviation from the radiation-dominated Universe as shown in
Fig.~\ref{fig:t_tH}.  Hence high frequency modes entering the horizon
before $\phi$-domination experience relative suppression compared with
low frequency modes.  As a result, as one can see, $\Omega_{\rm GW}$
is suppressed for high frequency modes which enter the horizon before
the $\phi$-domination.

In addition, most importantly, the effect of anisotropic stress caused
by dark radiation dissipates the GW energy density of the mode with
$k\lesssim k_{\rm dec}$,
because dark radiation is already created by the $\phi$ decay when such modes enter the horizon.
The effect is weaker for higher frequency because the abundance of $X$ is smaller at the horizon entry of high frequency modes.

Therefore, we expect suppression on the GW spectrum for both high frequency and low frequency sides :
the former caused by modified expansion rate due to $\phi$ and the latter by the anisotropic stress of $X$.
The GW spectrum between these two regimes, $k \sim k_{\rm dec}$, receives both effects and 
the resulting shape of the spectrum depends on how effective those effects are at $k \sim k_{\rm dec}$.
Numerical calculations show that a dip in the spectrum may appear at $k\sim k_{\rm dec}$.
In particular, the dip becomes more apparent
when $B_X$ is close to $1$.  Such a dip provides a smoking-gun
signature of the dark-radiation production by the decay of massive
particles.  If $\phi$ and $X$ are completely sequestered from the
standard-model sector, for example, $B_X=1$ may be realized.  Then,
such a model provides a striking signature in the GW spectrum.

Note that, in the low frequency limit $k\ll k_{\rm dec}$, we have numerically confirmed 
the suppression factor $C_2$ caused by dark radiation.
As a result, $\Omega_{\rm GW}/\Omega_{\rm GW}^{\rm (SM)}$ at $k\ll k_{\rm dec}$ is close to one 
as shown in Fig.~\ref{fig:norm}.

\section{Conclusions and Discussion}     \label{sec:conc}

In this paper we have studied the spectrum of inflationary GW background in the presence of dark radiation,
motivated by recent observational preferences for $\Delta N_{\rm eff}\sim 1$.
We have assumed that the dark radiation is non-thermally produced by decay of massive particles $\phi$.
There are several effects on the GW spectrum.
First, the equation of state of the Universe is modified due to the $\phi$ energy density
and it changes the shape of the GW spectrum.
Second, the anisotropic stress carried by dark radiation dissipates the GW amplitude
for modes entering the horizon around and after $\phi$ decay.
Numerical results show that there may appear a characteristic dip around $k\sim k_{\rm dec}$,
which is a smoking-gun signature of dark radiation.
It not only provides an evidence of dark radiation, but also sheds light on its production mechanism.

Some notes are in order.  We have assumed that the dark radiation
anisotropic stress is induced only by the primordial GWs.  This is not
in general true in the second order perturbation theory.
Free-streaming particles (as well as other fluids) contribute to GWs
at the second order in the scalar perturbation even if there is no
primordial tensor perturbation.  However, this contribution is
negligible for $r\gtrsim
10^{-6}$~\cite{Baumann:2007zm,Mangilli:2008bw}.

So far, we have considered dark radiation produced by the decay
of $\phi$.  However, it is possible that the dark radiation was once
in thermal equilibrium and decoupled from thermal bath at the
temperature $T_{\rm dec}$.  In this case, the extra effective number
of neutrino species is given by
\begin{equation}
  \Delta N_{\rm eff} = 
  \frac{4}{7}\epsilon N_X \left[ \frac{10.75}{g_{*s}(T_{\rm dec})} \right]^{4/3},
\end{equation}
where
\begin{equation}
  \epsilon = \begin{cases}
    1    & {\rm ~~for~a~real~scalar},\\
    7/4 & {\rm ~~for~a~chiral~fermion},
  \end{cases}
\end{equation}
and $N_X$ counts the number of $X$ species.  If the decoupling
temperature is higher than the weak scale, we need $N_X\gtrsim 20$ for
explaining $\Delta N_{\rm eff}\simeq 1$.  The modulation in the GW
spectrum, similar to the effect caused by of neutrinos apparent at the
GW frequency of $10^{-10}$\,Hz~\cite{Weinberg:2003ur,Watanabe:2006qe},
appears at the frequency inside the range of DECIGO/BBO sensitivities
for $T_{\rm dec}\sim 10^7$--$10^9$\,GeV.  If the decoupling
temperature is $\mathcal O$(1)\,MeV, $N_X\sim 1$ is sufficient in order
to obtain $\Delta N_{\rm eff}\simeq 1$ but the dip in the GW spectrum
cannot be seen in the GW detectors.  Instead, overall normalization of
the GW spectrum at the observable frequency range, inferred from the
measured tensor-to-scalar ratio, is enhanced by the factor $C_1\sim
1.3$.  (At this epoch, dark radiation took part in thermal bath and
there is no anisotropic stress damping on GW amplitudes with
corresponding modes.)  This provides another indirect evidence of dark
radiation.

\begin{acknowledgments}

This work is supported by Grant-in-Aid for
Scientific research from the Ministry of Education, Science, Sports,
and Culture (MEXT), Japan, No.\ 22540263 (T.M.), No.\ 22244021 (T.M.),
No.\ 23104001 (T.M.), No.\ 21111006 (K.N.), and No.\ 22244030 (K.N.).

\end{acknowledgments}

\appendix
\section{Equation of motion of gravitational waves with dark radiation}  

In this Appendix we derive Eq.~(\ref{eq_hdif_u}), the equation of
motion of GWs with dark radiation.  We follow
Refs.~\cite{Weinberg:2003ur,Watanabe:2006qe} but the result is
slightly different because $X$ is continuously produced by the decay
of $\phi$ so that the number of $X$ in the comoving volume is not
constant.

Throughout this appendix, we use the synchronous gauge and consider
tensor perturbations defined in \eqref[eq_line_element].

The equation of motion for tensor perturbations in Fourier space is
\begin{equation}
       \pardif[^2 h_{ij}][u^2] + 2 H_u \pardif[h_{ij}][u] + h_{ij} = 16 \pi G \left( \frac{a}{k} \right) ^2 \Pi_{ij},
    \label{eq_Einstein}
\end{equation}
where $u \equiv k \eta \equiv k \int_0^t \frac{dt'}{a(t')}$, $H_u \equiv \frac{1}{a} \pardif[a][u]$ and $\Pi_{ij}$ is defined by using the total energy momentum tensor as
\begin{eqnarray}
       T_{ij}^{\rm (tot)} &=& P g_{ij} + a^2 \Pi_{ij} , \\
       P &\equiv& \frac{1}{3} T^{i {\rm (tot)}}_{\; \;  i}.
       \label{eq_T}
\end{eqnarray}
Our goal in this appendix is to express the RHS of
  \eqref[eq_Einstein] in terms of metric perturbations.  In what
  follows, use the fact that only the collisionless particle (i.e.,
  $X$) contributes to the anisotropic stress $\Pi_{ij}$.

We first introduce the distribution function of the relativistic
  components $F^{\rm (tot)}(t,x^i,p_i)$, with which the total number
  of relativistic particles with particular momentum range contained
  in the volume element is given by $F^{\rm (tot)} dx^1 dx^2 dx^3 dp_1
  dp_2 dp_3$.  (Here and hereafter, $x^i$ is the comoving coordinate,
while $p_i$ is the comoving momentum.) Note that $F^{\rm (tot)}$ is a
scalar under general coordinate transformations which preserve the
synchronous gauge.  The distribution function can be decomposed as
\begin{eqnarray}
  F^{\rm (tot)}(t,x^i,p_i) = F^{(X)} (t,x^i,p_i)+ F^{\rm (rad)}(t,x^i,p_i),
\end{eqnarray}
where $F^{(X)}$ and $F^{\rm (rad)}$ are distribution functions of the dark
radiation $X$ and that of ordinary radiation (like photon, gluon, and
so on) with very short free-streaming length, respectively.
Hereafter, we omit the superscript $X$ for the distribution function of $X$ for notational simplicity :
$F (t,x^i,p_i) \equiv F^{(X)} (t,x^i,p_i)$.

We start with the effect of dark radiation on the anisotropic stress.
The distribution function of $X$ obeys the collisionless Boltzmann
equation with source from non-relativistic decaying particle
$\phi$:
\begin{equation}
       \frac{dF}{dt} 
       = \frac{B_X}{4 \pi (p^0)^3}\Gamma_\phi\rho_\phi
         \delta \left( p^0 - \frac{m_{\phi}}{2} \right),
  \label{eq_Boltz}
\end{equation}
where $p^0$ is the energy of $X$, and we assume that $\phi$ decays
into two $X$s.  Also note that $p^0$ and $p^{i}$ should be regarded as
functions of $p_i$ through $g_{\mu \nu} p^{\mu} p^{\nu} =
-(p^0)^2+a^2(\delta_{ij} + h_{ij}) p^i p^j = 0$ and $p^i = g^{ij} p_j
= a^{-2} (\delta_{ij} - h_{ij})p_j$. The LHS of Eq.~(\ref{eq_Boltz})
is
\begin{eqnarray}
  \begin{split}
    \frac{dF}{dt} 
    &= \frac{\partial F}{\partial t} + \frac{dx^i}{d t} \frac{\partial F}{\partial x^i} + \frac{dp_i}{d t} \frac{\partial F}{\partial p_i} \\
    &= \frac{\partial F}{\partial t} + \frac{p^i}{p^0} \frac{\partial F}{\partial x^i} 
    +\frac{1}{2} g_{ij,k} \frac{p^i p^j}{p^0} \frac{\partial F}{\partial p_k},
  \end{split}
  \label{eq_Boltz_lhs}
\end{eqnarray}
where we used
\begin{eqnarray}
       \frac{dx^i}{dt} 
       &=&
       \frac{p^i}{p^0} , \\
       \frac{dp_i}{dt}
       &=& \frac{1}{2} g_{jk,i} \frac{p^j p^k}{p^0}. \label{eq_geodesic}
      \end{eqnarray}
\eqref[eq_geodesic] is obtained from the geodesic equation.

Next we decompose $F$ into the unperturbed part $\bar{F}(t,p)$, where $p \equiv \sqrt{p_i p_i}$ should not be confused with the pressure, and the perturbed part $\delta F$. We further decompose $\delta F$ into two terms $\delta F_1$ and $\delta F_2$ for later convenience:
\begin{eqnarray}
       \delta F_1 (t,x^i,p_i) &\equiv& \bar{F} (t,(g^{ij} p_i p_j)^{1/2} / a) - \bar{F} (t,p) , \nonumber \\ \\
       \delta F_2 (t,x^i,p_i) &\equiv& F - \bar{F} - \delta F_1.
\end{eqnarray}
We get from Eq.~(\ref{eq_Boltz}) the zeroth-order equation
\begin{equation}
       \frac{\partial \bar{F}}{\partial t}
      = \frac{B_X}{4 \pi (\bar{p}^0)^3} \Gamma_\phi\rho_\phi
         \delta \left( \bar{p}^0 - \frac{m_{\phi}}{2} \right),
  \label{eq_zeroth}
\end{equation}
and the first-order one
\begin{eqnarray}
       \frac{\partial ( \delta F_1 + \delta F_2)}{\partial t}
       + \frac{\bar{p}^i}{\bar{p}^0} \frac{\partial (\delta F_1 + \delta F_2)}{\partial x^i}&& \nonumber \\
       + \frac{1}{2} (\delta g_{jk})_{,i} \frac{\bar{p}^j \bar{p}^k}{\bar{p}^0} \frac{\partial \bar{F}}{\partial p_i}
       = a\frac{\partial^2\bar F}{\partial p \partial t}\delta p^0.&&
  \label{eq_first}
\end{eqnarray}
\begin{figure}[t]
  \begin{center}
   \centerline{\epsfxsize=0.45\textwidth\epsfbox{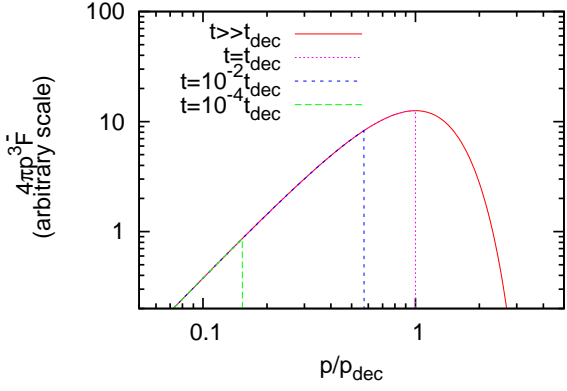}}
   \caption{$4 \pi p^3 \bar{F}$ as functions of $p$ (normalized by
     $p_{\rm dec} \equiv a(t_{\rm dec}) m_{\phi} / 2$).  Each line
     corresponds to $t \gg t_{\rm dec}$ (red solid), $t=10^{-4} t_{\rm dec}$
     (green long-dashed), $10^{-2} t_{\rm dec}$ (blue short-dotted) and
     $t_{\rm dec}$(magenta dotted).}
    \label{fig:p_F}
  \end{center}
\end{figure}

In \figref[fig:p_F], we show $4 \pi p^3 \bar{F}$ as a 
function of $p / p_{\rm dec}$, where $p_{\rm dec} \equiv a(t_{\rm dec}) m_{\phi}
/ 2$ is the comoving momentum of $X$ produced at $t=t_{\rm dec}$. We can see
that the energy fraction $4 \pi p^3 \bar{F}$ is mostly carried by $X$
produced at $t \simeq t_{\rm dec}$.  Then we use the following equations:
\begin{eqnarray}
      \delta F_1 &=& -\frac{1}{2} h_{ij} \frac{p_i p_j}{p} \pardif[\bar{F}][p], \label{eq_perturb_F1} \\
      \delta p^0 &=& -\frac{1}{2 a} \frac{h_{ij} p_i p_j}{p}, \label{eq_perturb_p^0} \\
      \delta p^i &=& -\frac{1}{a^2} h_{ij} p_j. \label{eq_perturb_p^i}
\end{eqnarray}
Using $\hat{p}_i \equiv p_i/p$ and substituting \eqref[eq_perturb_F1]
-- \eqref[eq_perturb_p^i] into \eqref[eq_first], we get
\begin{equation}
      \pardif[\delta F_2][t]
       + \frac{\hat{p}_i}{a} \frac{\partial \delta F_2}{\partial x^i}
       = \frac{1}{2} \pardif[h_{ij}][t] \pardif[\bar{F}][p] p \hat{p}_i \hat{p}_j.
\end{equation}
In terms of conformal time $\eta$, this equation is expressed as
\begin{equation}
      \pardif[\delta F_2][\eta]
       + \hat{p}_i \pardif[\delta F_2][x^i]
       = \frac{1}{2} \pardif[h_{ij}][\eta] \pardif[\bar{F}][p] p \hat{p}_i \hat{p}_j.
\end{equation}
In Fourier space,
\begin{equation}
       \pardif[\delta F_2][\eta]
       + i k \mu \delta F_2
       = \frac{1}{2} \pardif[h_{ij}][\eta] \pardif[\bar{F}][p] p \hat{p}_i \hat{p}_j,
    \label{eq_deq_deltaF2}
\end{equation}
where 
\begin{eqnarray}
       \delta F_2(\eta,x^i,p_i)
       &=& \int \frac{d^3 k}{(2 \pi)^3} \delta F_2(\eta,k_i,p_i) e^{i k_i x^i}, \\
       h_{ij} (\eta,x^i)
       &=& \int \frac{d^3 k}{(2 \pi)^3} h_{ij} (\eta,k_i) e^{i k_i x^i}, \\
       \mu
       &\equiv& \hat{k}_i \hat{p}_i.
\end{eqnarray}
We can use line-of-sight integral to get the solution of \eqref[eq_deq_deltaF2]:
\begin{equation}
       \delta F_2 = \int_0^{\eta} d\eta' \frac{1}{2} \pardif[h_{ij}][\eta] (\eta') \pardif[\bar{F}][p] (\eta') p \hat{p}_i \hat{p}_j e^{-ik \mu (\eta-\eta')},
 \label{eq_deltaF2}
\end{equation}
where we have used $\delta F_2(\eta=0)=0$ because there is no $X$ in
the beginning.

We take the first-order perturbation of the energy-momentum tensor of
$X$:
\begin{eqnarray}
  T_{\mu \nu}^{(X)}
  &=& \frac{1}{\sqrt{-{\rm det} g_{\mu \nu}}} 
  \int d^3p F \frac{p_{\mu} p_{\nu}}{p^0}, \\
  \delta T_{ij}^{(X)}
  &=& \frac{1}{a^3} \int d^3p \left[ (\delta F_1 +\delta F_2) 
    \frac{p_i p_j}{\bar{p}^0}
    +  \bar{F} p_i p_j \delta \left( \frac{1}{p^0} \right) \right]. 
  \nonumber \\
\end{eqnarray}
Note that energy momentum tensor defined above transforms as a tensor under
general coordinate transformations since $\int d^3p / p^0 \propto d^4p
\delta (g^{\mu \nu} p_{\mu} p_{\nu} )$.  Using \eqref[eq_perturb_F1]
-- \eqref[eq_perturb_p^i], we get
\begin{eqnarray}
  \delta T_{ij}^{(X)}
  &=&
  \frac{1}{a^3} \int d^3p \left[ \frac{}{} \delta F_2 a p \hat{p}_i \hat{p}_j \right. \nonumber\\
  &\left. \right.& \left. - \frac{1}{2} a h_{kl} p^2 \hat{p}_i \hat{p}_j \hat{p}_k \hat{p}_l \pardif[\bar{F}][p]
    + \frac{1}{2} a h_{kl} p \hat{p}_i \hat{p}_j \hat{p}_k \hat{p}_l \bar{F} \right] \nonumber \\
  &=&
  \frac{1}{a^3} \int d^3p \delta F_2 a p \hat{p}_i \hat{p}_j \nonumber \\
  &\left. \right.& + \frac{1}{a^3} \int dp p^2  \left[ -\frac{1}{2} a h_{kl} p^2 \pardif[\bar{F}][p] 
    + \frac{1}{2} a h_{kl} p \bar{F} \right] \nonumber \\
  &\left. \right.& \times \frac{4 \pi}{15} (\delta_{ij} \delta_{kl} + \delta_{ik} \delta_{jl} + \delta_{il} \delta_{jk} ) \nonumber \\
  &=&
  \frac{1}{a^2} \int d^3p \delta F_2 p \hat{p}_i \hat{p}_j + \frac{1}{3} a^2 h_{ij} \rho_X.
  \label{eq_deltaT_1}
\end{eqnarray}
Here, we used
\begin{eqnarray}
  &\left. \right.&
  \int d \Omega_p \hat{p}_i \hat{p}_j \hat{p}_k \hat{p}_l e^{-i \hat{p}_i \hat{k}_i u} \nonumber \\
  &\left. \right.&
  = 4 \pi \left[ j_4 (u) \hat{k}_i \hat{k}_j \hat{k}_k \hat{k}_l - \frac{j_3 (u)}{u} (\hat{k}_i \hat{k}_j \delta_{kl} + {\rm 5~perms}) \right. \nonumber \\
  &\left. \right.& \; \; \;
  + \left. \frac{j_2 (u)}{u^2} (\delta_{ij} \delta_{kl} + {\rm 2~perms}) \right]
    \label{eq_solid_integral}
\end{eqnarray}
and
\begin{equation}
  \int dp 4 \pi p^3 \bar{F} = a^4 \rho_X,
\end{equation}
where $\rho_X$ is the energy density of $X$ and $j_n$ is the $n$-th
spherical Bessel function.

Next, we consider the effect of $F^{\rm (rad)}$, for which $\delta
F_2^{\rm (rad)}=0$ because the free-streaming length is very short.
Then, we obtain
\begin{eqnarray}
  \delta T_{ij}^{\rm (rad)} = \frac{1}{3} a^2 h_{ij} \rho_{\rm rad}.
  \label{eq_deltaTrad}
\end{eqnarray}
We also note that perturbation in the energy momentum tensor of $\phi$ vanishes since it behaves as non-relativistic matter :
\begin{equation}
  \delta T_{ij}^{\rm (\phi)} = 0.
   \label{eq_deltaTphi}
\end{equation}

Taking the first-order perturbation of \eqref[eq_T], we obtain
\begin{eqnarray}
  \delta T_{ij}^{\rm (tot)}
  &=& \delta P \cdot \bar{g}_{ij} + \bar{P} \cdot \delta g_{ij} + a^2 \Pi_{ij} \nonumber \\
  &=& \frac{1}{3} a^2 h_{ij}(\rho_X+\rho_{\rm rad}) + a^2 \Pi_{ij},
  \label{eq_deltaT_2}
\end{eqnarray}
where we used \eqref[eq_perturb_F1] -- \eqref[eq_perturb_p^i],
\eqref[eq_deltaF2], \eqref[eq_solid_integral], $h_{ii} = 0$ and
$\delta P=0$.  The last condition comes from the fact that tensor
perturbations cannot produce perturbations in scalar variables. 
Using \eqref[eq_deltaT_1], \eqref[eq_deltaTrad], \eqref[eq_deltaTphi], and \eqref[eq_deltaT_2], we
obtain
\begin{equation}
       a^2 \Pi_{ij} =  \frac{1}{a^2} \int d^3p \delta F_2 p \hat{p}_i \hat{p}_j.
    \label{eq_Pi}
\end{equation}


Substituting \eqref[eq_Pi] into the RHS of \eqref[eq_Einstein], we
obtain
\begin{eqnarray}
  &\left. \right.&
  \pardif[^2 h_{ij}][u^2] + 2 H_u \pardif[h_{ij}][u] + h_{ij} \nonumber \\
  &\left. \right.&
  = 16 \pi G \left( \frac{a}{k} \right)^2 \frac{1}{a^4} \int d^3p p \hat{p}_i \hat{p}_j \nonumber \\
  &\left. \right.& \; \; \;
  \times \int_0^{\eta} d\eta' \frac{1}{2} \pardif[h_{kl}][\eta] (\eta') \pardif[\bar{F}][p] (\eta') p \hat{p}_k \hat{p}_l
  e^{-ik \mu (\eta-\eta')} \nonumber \\
  &\left. \right.&
  = -8 \pi G \frac{1}{k^2 a^2} \int _0^{\eta} d\eta' \pardif[h_{kl}][\eta] (\eta') \nonumber \\
  &\left. \right.& \; \; \; 
  \int d\Omega_p \frac{1}{\pi} a^4 \rho_X (\eta') \hat{p}_i  \hat{p}_j \hat{p}_k \hat{p}_l  e^{-ik \mu (\eta-\eta')} \nonumber \\
  &\left. \right.&
  =-24 H_{u}^2 \frac{1}{a^4 \rho_{\rm tot} (u)} \int _0^{u} du'
  a^4 \rho_X (u') \pardif[h_{ij}][u] (u') \frac{j_2 (u - u')}{(u - u')^2}, \nonumber \\
\end{eqnarray}
where we used partial integration, \eqref[eq_solid_integral] and Friedmann equation $H_{u}^2 = 8 \pi G \rho_{\rm tot} a^2 / 3 k^2$. After decomposing $h_{ij}$ using \eqref[eq_h_decompose], we finally obtain \eqref[eq_hdif_u].



\end{document}